\documentclass[conference]{IEEEtran}
\IEEEoverridecommandlockouts
\usepackage{amsmath,amssymb,amsfonts}
\usepackage{algorithmic}
\usepackage{graphicx}
\usepackage[utf8]{inputenc}    
\usepackage{hyperref}
\usepackage[maxbibnames=6]{biblatex}
\def\BibTeX{{\rm B\kern-.05em{\sc i\kern-.025em b}\kern-.08em
    T\kern-.1667em\lower.7ex\hbox{E}\kern-.125emX}}
\begin{document}

\title{Practical and Verifiable Electronic Sortition}

\author{\IEEEauthorblockN{Hsun Lee}
\IEEEauthorblockA{\textit{Computer Science and Information Engineering} \\
\textit{National Taiwan University}\\
Taipei, Taiwan \\
leexun@csie.ntu.edu.tw}
\and
\IEEEauthorblockN{Hsu-Chun Hsiao}
\IEEEauthorblockA{\textit{Computer Science and Information Engineering} \\
\textit{National Taiwan University}\\
Taipei, Taiwan \\
hchsiao@csie.ntu.edu.tw}
}

\maketitle

\begin{abstract}

Existing verifiable e-sortition systems are impractical due to computationally expensive verification (linear to the duration of the registration phase, T) or the ease of being denial of service. Based on the advance in verifiable delay functions, we propose a verifiable e-sortition scheme whose result can be efficiently verified in constant time with respect to T. We present the preliminary design and implementation, and explore future directions to further enhance practicability.
\end{abstract}

\begin{IEEEkeywords}
verifiable sortition, verifiable delay functions
\end{IEEEkeywords}

\section{Introduction}
Electronic sortition (e-sortition) plays a principal role in democratic societies. It enables fair distribution of limited resources, such as the right to rent social housing~\cite{chang2013public} or to pre-order masks~\cite{emask2020}~\cite{10.1001/jama.2020.3151} during epidemic prevention.
A typical e-sortition system randomly selects a subset from the registered users by using a centralized server. To reduce trust in the centralized server, it is desirable to have \emph{verifiable e-sortition}, whose fairness can be publicly verified.

However, existing verifiable e-sortition schemes are impractical due to computationally expensive verification or the ease of being denial of service.
For example, some systems~\cite{chow2006practical,liu2006new} apply \emph{delay functions}, a kind of moderately hard cryptographic functions, to prevent malicious users from manipulating the result before the end of the sortition registration. These systems require verifiers to re-compute the delay function, which takes longer than the time of the registration phase (e.g.,several days).
Some systems~\cite{he2009electronic,ramezanian2018decentralized} rely on commit-reveal protocols to prevent such manipulation, but they are vulnerable to denial of service attacks caused by participants withholding the confirmation messages.

In order to construct a practical verifiable e-sortition system, we adopt the \emph{verifiable delay function (VDF)}, which is a special type of delay functions whose output can be efficiently verified~\cite{boneh2018verifiable}. Our scheme can securely generate unpredictable and unbiased pseudo-random result without any trusted third party, and improve the time complexity of delay function verification from linear to constant.

\subsection{The scheme}
\label{the_scheme}
Throughout this paper, $x_u$ denotes a string generated by a user $u$, $y$ denotes the verifiable random output (which will seed a public pseudo-random number generator for e-sortition), and $\pi$ denotes the proof of the output correctness. $T$ is a public parameter indicating the computation time, which must be slightly longer than the time range of the registration. $N$ is the final number of registered users. Furthermore, our scheme can be initialized without any trusted third party, because we apply the implementation of VDF proposed by Wesolowski~\cite{wesolowski2019efficient}. Our scheme can be divided into three phases:
\begin{enumerate}
  \item \textbf{Registration.} Each user $u$ selects and sends a arbitrary $x_u$ to the server. The server receives $x_u$ and returns a digitally signed confirmation as a proof of registration.
  \item \textbf{Result generation.} The server builds a Merkle tree with all $x_u$s as the leaves, and obtains the root of the Merkle tree, $x_{root}$. The server then evaluates the VDF function $VDFEval(x_{root}, T) = (y, \pi)$. After computation, the server seeds a public PRNG using $y$ to select the winners. Finally, the server publishes $(y, \pi)$ to all users.
  \item \textbf{Verification.} Every user $u$ can verify whether his or her $x_u$ was correctly included during result generation. Specifically, the user requests the server for the Merkle audit path of $x_u$ and re-computes the root hash of the Merkle tree to verify the inclusion proof. After completion, the user computes $VDFVerify(x_{root}, y, \pi, T) = \{Yes,No\}$ to determine the correctness of $(y, \pi)$. Notice that $VDFVerify$ is much faster than $VDFEval$, because it is not a re-computation of $VDFEval$. If any of the verification fails, the user can prove the sortition is invalid by providing the information above.
\end{enumerate}


\section{Result}
Compared with prior work, our construction reduces the time and memory complexity of the verification phase by applying the VDF and Merkle tree (Table \ref{tab:complexity}). In addition, we decrease the maximum time of $H_{prime}$ function in VDF verification in web browsers from 10.7s to 683ms by skipping unnecessary procedures of primality tests (Appendix~\ref{experiment}).




\begin{table}[htbp]
\vspace{-3mm}
\begin{center}
\rule{0pt}{1ex}
\caption{Complexity Improvement}
\begin{tabular}{|c|c|c|}
\hline
\textbf{\textit{}}& \textbf{\textit{Verification of delay function}}& \textbf{\textit{Size of inclusion proof}} \\
\hline
[1] & $\mathcal{O}(T)$ & $\mathcal{O}(N)$ \\
\hline
[2] & $\mathcal{O}(T)$ & $\mathcal{O}(\log{(N)})$ \\
\hline
Our work & $\mathcal{O}(1)$ & $\mathcal{O}(\log{(N)})$ \\
\hline


\end{tabular}
\label{tab:complexity}
\end{center}
\vspace{-3mm}
\end{table}


For future work, we are extending our scheme to provide a verifiable e-sortition service that abstracts the implementation of VDFs. We envision that such a loosely coupled service can allow existing non-verifiable e-sortition systems to rapidly transit to verifiable ones.




\begin{figure}[t]
  \centering
    \includegraphics[scale=0.35]{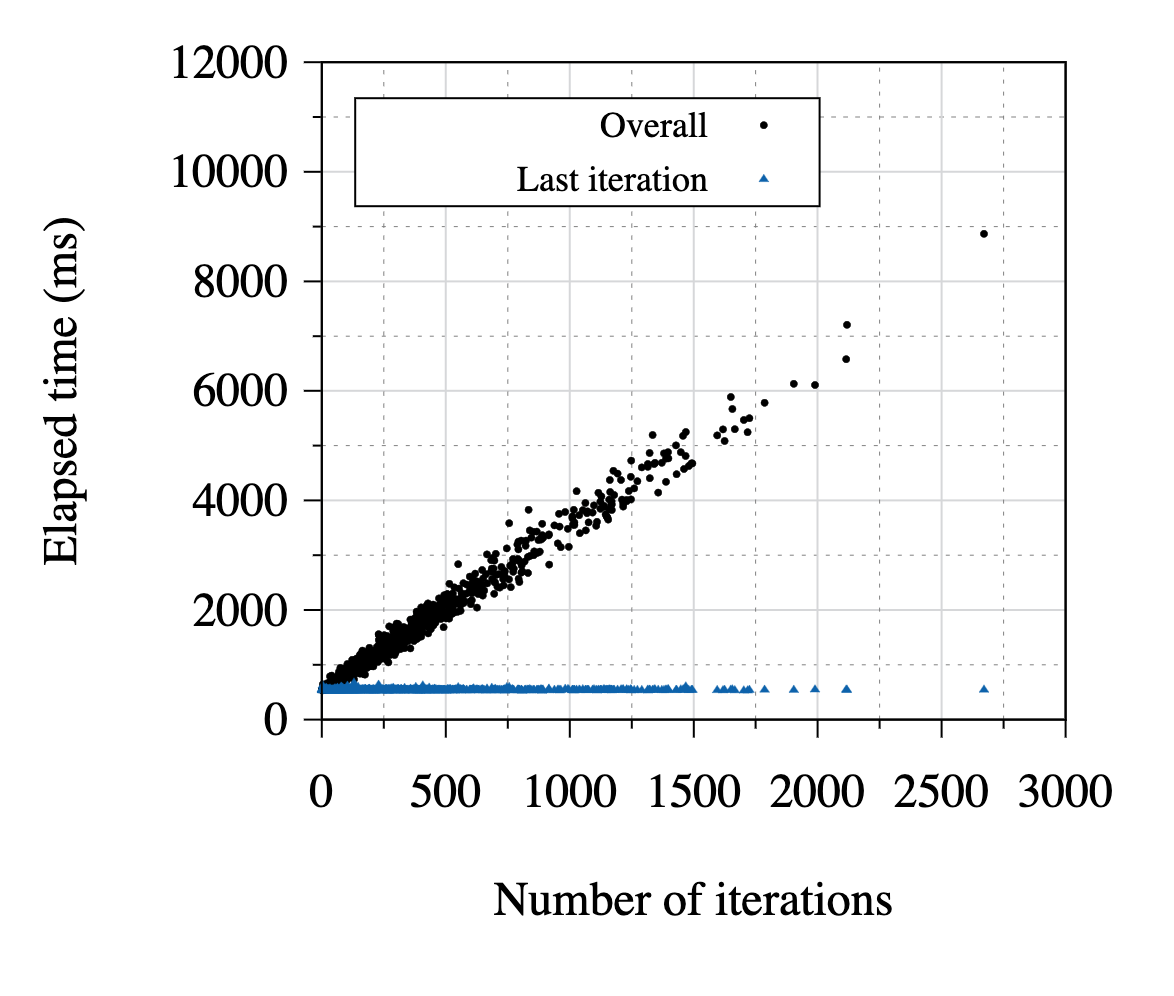}
    \vspace{-5mm}
    \caption{$H_{prime}$ in Safari 12.1 on MacBook Pro with 2.7 GHz Intel Core i5}
  \label{MacBookPro}
  \vspace{-3mm}
\end{figure}

\textbf{Acknowledgements.} This research was supported by the Ministry of Science and Technology of Taiwan under grant MOST 109-2636-E-002-021.

\printbibliography

\appendices


\section{EXPERIMENT RESULT}
\label{experiment}

\begin{figure}[t]
  \centering
    \includegraphics[scale=0.35]{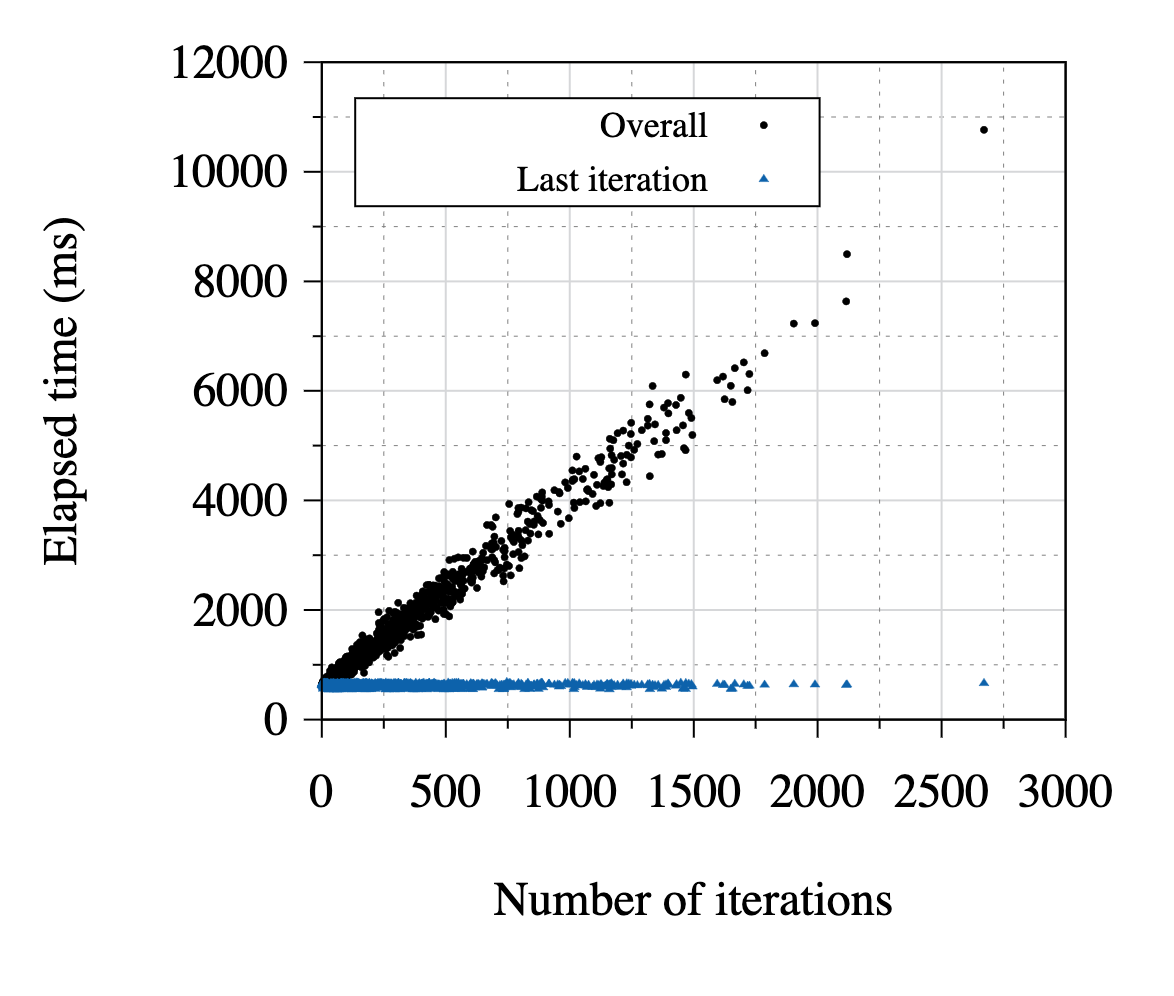}
    \vspace{-5mm}
    \caption{$H_{prime}$ in Safari 12.1 on iPhone 6 with Dual-core 1.4 GHz Typhoon}
  \label{iPhone6}
  \vspace{-3mm}
\end{figure}

We conduct preliminary experiments to evaluate the verification speed in web browsers on laptop (Fig.~\ref{MacBookPro}) and mobile phone (Fig.~\ref{iPhone6}). We use the VDF implemented by Chai Network~\cite{chaivdf}, because Chia opened a competition for the fastest VDF to precisely evaluate the security of the Time-lock assumption in VDF~\cite{wesolowski2019efficient}. Additionally, we modify the implementation of the VDF to cross-compile it from C++ into WebAssembly to ensure future portability. 

However, we expect there will be a performance downgrade after moving native program into browser. For example, the procedure of primality test. The VDF proposed by Wesolowski~\cite{wesolowski2019efficient} requires the server and the users themselves to compute the $x_{root}$ into a negative prime $d$ of large absolute value, and such that $d \equiv 1 \pmod{4}$. This procedure involves $H_{prime}(s)=p$, a deterministic algorithm takes arbitrary string $s$, iteratively performs hash and primality test, and finally returns a large psuedoprime p ($2^{1023} \leq p < 2^{1024}$). It might take lots of iterations to complete and thus increases potential overhead.

To estimate the overhead of primality test, we generate 1024 sample strings as $s$ to measure the elapsed time of $H_{prime}$ in web browsers. We use \emph{mpz\_probab\_prime\_p} in GMP 6.2.0 with 50 repetitions to perform the test. As shown in Fig.~\ref{MacBookPro}~and~\ref{iPhone6}, it takes around 547ms to pass the final primality test, which occurs in every final iteration before $H_{prime}$ returns.  Namely, most of time in $H_{prime}$ is spent on those failed primality tests.

We propose a solution to reduce this overhead. The server can publish a value $i$, the number of the iterations to find the psuedoprime, with $(y,\pi)$ after result generation. Therefore, users can perform $i-1$ iterations and only do the primality test in the $i$th iteration to ensure the final $d$ is truly a psuedoprime. By this method, the maximum time of $H_{prime}$ in our experiment on iPhone 6 can be decreased from 10.7s to 683ms. The data of these experiments can be found at \href{https://github.com/leexun/verifiable-sortition-system}{https://github.com/leexun/verifiable-sortition-system}.





\end{document}